# Electrical properties of isotopically enriched neutron-transmutation-doped $^{70}$Ge:Ga near the metal-insulator transition

Michio Watanabe and Youiti Ootuka

*Cryogenic Center, University of Tokyo, 2-11-16 Yayoi, Bunkyo-ku, Tokyo 113-0032, Japan*

Kohei M. Itoh

*Department of Applied Physics and Physico-Informatics, Keio University, 3-14-1 Hiyoshi, Kohoku-ku, Yokohama 223-8522, Japan*

Eugene E. Haller

*Lawrence Berkeley National Laboratory and University of California at Berkeley, Berkeley, California 94720*

We report low-temperature carrier transport properties of a series of nominally uncompensated neutron-transmutation-doped $^{70}$Ge:Ga samples very close to the critical concentration $N_c$ for the metal-insulator transition. The nine samples closest to $N_c$ have Ga concentrations $N$ in the range $0.99N_c < N < 1.01N_c$. The electrical conductivity $\sigma$ has been measured in the temperature range $T = 0.02 - 1$ K. On the metallic side of the transition the standard $\sigma(T) = a + bT^q$ with $q = 1/2$ was observed for all the samples except for the two that are closest to $N_c$ with $N$ between $N_c$ and $1.0015N_c$. These samples clearly show $q = 1/3$.

An extrapolation technique has been developed in order to obtain the zero-temperature conductivity $\sigma(0)$ from $\sigma(T)$ with different dependence on $T$. Based on the analysis, $\nu \approx 0.5$ in the familiar form of $\sigma(0) \propto (N/N_c - 1)^\nu$ has been found. On the insulating side of the transition, variable range hopping resistivity $\rho(T) \propto \exp(T_0/T)^p$ with $p = 1/2$ has been observed for all the samples having $N < 0.991N_c$. In this regime $T_0 \propto (1 - N/N_c)^\alpha$ with $\alpha \approx 1$ as $N \to N_c$. The values of $T_0$ agree very well with theoretical estimates based on the modified Efros and Shklovskii relation $k_B T_0 \approx (2.8 e^2/4\pi\epsilon_0\kappa_0\xi_0)(1 - N/N_c)^\alpha$, where $\kappa_0$ and $\xi_0$ are the dielectric constant and the Bohr radius, respectively. The insulating samples very close to the transition ($0.991N_c < N < N_c$) exhibit quite a different behavior. In this range $1/p$ increases rapidly as $N$ changes from $0.991N_c$ to $N_c$. The relevance of our findings to the collapsing of the Coulomb gap is discussed.

## I. INTRODUCTION

The doping-induced metal-insulator (MI) transition in semiconductors has been studied extensively in the past few decades.[1–3] However, there still remain a number of major theoretical and experimental challenges. Measurements of the electrical conductivity $\sigma(T)$ as a function of temperature near the MI transition are fundamental to the understanding of the roles of potential disorder and electron-electron interaction. The zero-temperature conductivity $\sigma(0)$ obtained from an appropriate extrapolation of the temperature dependent $\sigma(T)$ to zero temperature is evaluated as a function of doping concentration $N$ immediately above the critical concentration $N_c$ for the MI transition;

$$\sigma(0) = \sigma_0 (N/N_c - 1)^\nu, \qquad (1)$$

where $\sigma_0$ is the prefactor and $\nu$ is the critical exponent. In several strongly disordered systems, e.g., compensated single crystalline semiconductors [Ge:Sb,[5] Si:P,B,[6] Ge:Ga,As,[7] Al$_{0.3}$Ga$_{0.7}$As (Ref. 8)] and amorphous alloys,[9–11] $\nu \approx 1$ has been found. These results are in good agreement with the prediction[4] for the transition driven by disorder. It was also found that in compensated Al$_{0.3}$Ga$_{0.7}$As the dielectric constant on the insulating side diverges with the critical exponent of $s \approx 2.3$ near the transition,[8] i.e., $s \approx 2\nu$ predicted[12] for the disorder-induced transition holds. Thus there is strong evidence that the effect of disorder rather than electron-electron interaction plays the key role in the MI transition of compensated semiconductors. On the other hand, a critical exponent of $\nu \approx 0.5$ has been obtained with a number of nominally uncompensated semiconductors [Si:P,[13,14] Si:As,[15,16] Ge:As,[17] Si:B,[18] Ge:Ga (Ref. 19)]. This value of $\nu \approx 0.5$ is significantly smaller than $\nu \approx 1 - 1.3$ predicted[4,20–24] by the transition purely driven by the disorder. It also does not satisfy Chayes *et al.*'s inequality[25] $\nu > 2/3$ for transitions due to both disorder and electron-electron interaction. In response to these discrepancies, several theoretical ideas supporting $\nu \approx 0.5$ have been proposed.[26–28] However, general agreement between the experimental results and theory has yet to be achieved by any of the models.[26–28] The interesting observation reported commonly on uncompensated systems is the relatively wide range of $N$ above $N_c$ (typically up to $1.5 N_c$ or larger) in which $\sigma(0)$ can be fitted very well with Eq. (1) with $\nu \approx 0.5$. Based on this observation, Fritzsche[29] proposed a model composed of one main transition accompanied by two satellite transitions, one on each side of $N_c$. Stupp et al.[30] questioned the large critical region and found a narrow regime $N_c < N < 1.1 N_c$ in which $\sigma(0)$ of uncompensated Si:P exhibits $\nu \approx 1.3$. More recently, $\nu \approx 1$ was claimed also for uncompensated Ge:As by Shlimak et al.[31] This recent trend of $\nu$ moving from $\approx 0.5$ to $\approx 1 - 1.3$ was ended by our work on homogeneously doped, nominally uncompensated Ge:Ga, in which $\nu \approx 0.5$ was established unambiguously.[19] The exponents $\nu \approx 1 - 1.3$ claimed for melt-doped Si:P (Ref. 30) and Ge:As (Ref. 31) should be interpreted with great caution for the reasons we give in the following paragraphs.

In the experiment reported here we probe the low-temperature electrical properties of nominally uncompensated Ge:Ga in the region extremely close to the MI transition; $0.99 N_c < N < 1.01 N_c$. This concentration regime has not been fully investigated in our earlier work.[19] For the case of melt- (or metallurgically) doped samples that have been employed in most of the previous studies,[13–18,30,31] the spatial fluctuation of $N$ due to dopant striations and segregation can easily be on the order of 1% across a typical sample for the four-point resistance measurement (length of $\sim 5$ mm or larger).[32] For this reason it is not meaningful to discuss physical properties in this truly critical region (e.g., $|N/N_c - 1| < 1\%$) based on the data taken with melt-doped samples.

A precise determination of $N$ in a melt-doped sample is also difficult due to the spatial fluctuation of $N$ as well as to the limited accuracy of the existing method to measure $N$ near the transition. The determination of $N$ by Hall effect may be inaccurate due to the possible divergence of the Hall coefficient from unity near the transition. Resistivity measurements at two temperatures (4.2 K and 300 K) (Ref. 15) to find $N$ require an accurate calibration that cannot be established easily.

All Ge:Ga samples used in this work (and in our earlier study[19]) were prepared by neutron-transmutation doping (NTD) of isotopically enriched $^{70}$Ge single crystals. Our NTD method inherently guarantees the random distribution of the dopants down to the atomic level.[33–35] The $N$ for each sample is given by the thermal neutron fluence and its relation to $N$ has been accurately established[19] for $^{70}$Ge. We prepared 13 new NTD $^{70}$Ge:Ga samples with nine of them in the $0.99 N_c < N < 1.01 N_c$ region. The sample with the $N$ closest to $N_c$ has $N = 1.0004 N_c$. To our knowledge, neither experimental nor numerical studies on the MI transition have ever approached $N_c$ as close as this work has. Our study focuses on the analysis of the temperature dependence of $\sigma(T)$ below 1 K on both sides of the transition; the insulating phase ($N < N_c$) and the metallic phase ($N > N_c$). We investigate the universality of the $\sigma(T)$ in the metallic phase by introducing a numerical procedure. A quantitative discussion of $\sigma(T)$ in the insulating phase will be given in the context of the variable range hopping conduction model.

## II. EXPERIMENT

### A. Sample preparation and characterization

We first describe the preparation of the neutron-transmutation-doped $^{70}$Ge:Ga samples for the low-temperature conductivity measurements in the critical regime of the MI transition. We use NTD since it is known to produce the most homogeneous, perfectly random dopant distribution down to the atomic level.[33–35] The Czochralski grown, chemically very pure $^{70}$Ge crystal has isotopic composition $[^{70}$Ge$]=96.2$ at. % and $[^{72}$Ge$]=3.8$ at. %. The as-grown crystal is free of dislocations, $p$ type with an electrically active net-impurity concentration less than $5 \times 10^{11}$ cm$^{-3}$. The thermal neutron irradiation leading to NTD was performed at the University of Missouri Research Reactor with the thermal to fast neutron ratio of $\sim 30:1$. Upon capturing a thermal neutron $^{70}$Ge becomes $^{71}$Ge which decays with a half-life of 11.2 days via electron capture to a $^{71}$Ga acceptor. The small fraction of $^{72}$Ge becomes $^{73}$Ge which is stable, i.e., no further acceptors or donors are introduced. The post NTD rapid-thermal annealing at 650 °C for 10 sec removed most of the irradiation-induced defects from the samples. The short annealing time is important in order to avoid the redistribution and/or clustering of the uniformly dispersed $^{71}$Ga acceptors. The concentration of the electrically active radiation defects measured with deep level transient spectrometry (DLTS) after the annealing is less than 0.1% of the Ga concetration,[36] i.e., the compensation ratio of the samples is less than 0.001.[37] The dimension of most samples used for conductivity measurements was $6 \times 0.9 \times 0.7$ mm$^3$. Four strips of boron-ion-implanted regions on a $6 \times 0.9$ mm$^2$ face of each sample were coated with 200 nm Pd and 400 nm Au pads using a sputtering technique. Annealing at 300 °C for one hour activated the implanted boron and removed the stress in the metal films.

The Ga concentration $N$ in our $^{70}$Ge samples after NTD is given precisely by

$$[^{71}\text{Ga}] \ (\text{cm}^{-3}) = 0.1155 \times n \ (\text{cm}^{-2}), \tag{2}$$

where $n$ is the thermal neutron fluence.[37] The main goal of this study was to fill the gap in $N$ between 1.840 and $1.861 \times 10^{17}$ cm$^{-3}$ that was missing in our earlier work,[19] i.e., a precise control of $n$ of the order of 0.1% is needed. Although obtaining such a precision in $n$ seems difficult, we successfully used the following approach. When we prepared the insulating samples for our previous study,[19] we doped three 2 cm diameter wafers to $N = 1.733 \times 10^{17}$ cm$^{-3}$. For the present study 13 pieces were cut from two of the $N = 1.733 \times 10^{17}$ cm$^{-3}$ wafers. Each of the 13 pieces were then irradiated a second time to cover the range $N = 1.840 - 1.861 \times 10^{17}$ cm$^{-3}$ with a neutron fluence resolution of $n = 2.2 \times 10^{15}$ cm$^{-2}$ which corresponds to $N = 2.5 \times 10^{14}$ cm$^{-3}$ according to Eq. (2).

### B. Measurements

The electrical conductivity measurements were carried out down to temperatures of 20 mK using a $^3$He-$^4$He dilution refrigerator. All the electrical leads were low-pass filtered at the top of the cryostat. The sample was fixed in the mixing chamber and a ruthenium oxide thermometer [Scientific Instrument (SI), RO600A, 1.4×1.3×0.5 mm$^3$] was placed close to the sample. To measure the resistance of the thermometer, we used an ac resistance bridge (RV-Elekroniikka, AVS-47). The thermometer was calibrated against 2Ce(NO$_3$)$_3 \cdot$3Mg(NO$_3$)$_2 \cdot$24H$_2$O (CMN) susceptibility and against the resistance of a canned ruthenium oxide thermometer (SI, RO600A2) which was calibrated commercially over a temperature range from 50 mK to 20 K. We employed an ac method at 21.0 Hz to measure the resistance of the sample. The power dissipation was kept below $10^{-14}$ W, which is small enough to avoid overheating of the samples. The output voltage of the sample was detected by a lock-in amplifier (EG&G Princeton Applied Research, 124A). All the analog instruments as well as the cryostat were placed inside a shielded room. The output of the instruments was detected by digital voltmeters placed outside the shielded room. All the electrical leads into the shielded room were low-pass filtered. The output of the voltmeters was read by a personal computer via GP-IB interface connected through an optical fiber.

### III. RESULTS AND DISCUSSION

#### A. Electrical transport and the critical conductivity exponent in the metallic samples

The temperature dependence of the electrical conductivity mostly for the metallic samples is shown in Fig. 1. The solid symbols denote the data taken with the samples prepared in this work and the open ones are the data reevaluated with most of the samples described in Ref. 19. Note that several samples are doped successfully in the immediate vicinity of $N_c$. Mott's minimum metallic conductivity $\sigma_{\min}$ for Ge:Ga is estimated to be 7 S/cm using the relation $\sigma_{\min} \equiv C_M(e^2/\hbar)N_c^{1/3}$ with $C_M \approx 1/20$ as it was used for Si:P.[13,14] Figure 1 clearly shows that $\sigma$ of some of the newly prepared metallic samples takes values less than $\sigma_{\min}$ even at finite temperatures. The critical exponent $\nu$ in Eq. (1) is defined for the critical region $(N/N_c - 1) \ll 1$ through the conductivity at zero temperature $\sigma(0)$. Experimentally, however, it is impossible to reach $T = 0$ and a suitable extrapolation is required.

The temperature variation of the conductivity is governed mainly by the electron-electron interaction and can be written as

$$\Delta\sigma(T) \equiv \sigma(T) - \sigma(0) = m\sqrt{T}, \qquad (3)$$

where

$$m = A/\sqrt{D}. \qquad (4)$$

Here, $A$ is a temperature independent constant and $D$ is the diffusion constant, which is related to the conductivity via the Einstein relation

$$\sigma = (\partial n/\partial \mu)e^2 D, \qquad (5)$$

where $(\partial n/\partial \mu)$ is the density of states at the Fermi level. In the limit of $\Delta\sigma(T) \ll \sigma(0) \approx \sigma(T)$, $D$ can be considered as a constant, i.e., $m$ is constant. Usually $\sigma(0)$ is obtained by extrapolating $\sigma(T)$ to $T = 0$ assuming $\sqrt{T}$ dependence based on Eq. (3). Such an analysis was performed in our earlier work since $\Delta\sigma(T) \propto \sqrt{T}$ was found for all the samples.[19] It should be pointed out, however, that the above inequality $\Delta\sigma(T) \ll \sigma(0)$ is no longer valid as $N$ approaches $N_c$ from the metallic side since $\sigma(0)$ also approaches zero. In such cases $m$ in Eq. (3) is not temperature independent and $\Delta\sigma(T)$ may exhibit a temperature dependence different from $\sqrt{T}$. To examine this point for our experimental results, we go back to Fig. 1. It is seen here that the $\Delta\sigma(T)$ of the bottom five curves are not proportional to $\sqrt{T}$ while $\Delta\sigma(T)$ of the other higher $N$ samples are well described by $\propto \sqrt{T}$. The close-ups of $\sigma(T)$ for the six samples with positive $d\sigma/dT$ in the scale of $\sqrt{T}$ and $T^{1/3}$ are shown in Figs. 2(a) and 2(b), respectively. The upper and lower dotted lines represent the best fit using the data between 0.05 K and 0.5 K for the samples with $N = 1.912 \times 10^{17}$ cm$^{-3}$ and $N = 1.861 \times 10^{17}$ cm$^{-3}$, respectively. Each fit is shifted downward slightly for easier comparison. From this comparison it is clear that a $T^{1/3}$ dependence holds for samples in the very vicinity of the MI transition. The opposite is true for the curve at the top. This means that the $\sqrt{T}$ dependence in Eq. (3) is replaced by a $T^{1/3}$ dependence as the MI transition is approached.

A $T^{1/3}$ dependence close to the MI transition was predicted originally by Al'tshuler and Aronov.[38] They considered an interacting electron system with paramagnetic impurities, for which they obtained a single parameter scaling equation. At finite temperatures, they assumed a scaling form for conductivity according to the scaling hypothesis;

$$\sigma = \frac{e^2}{\hbar \xi} f(\xi/L_T), \qquad (6)$$

where $\xi$ is the correlation length and $L_T \equiv \sqrt{\hbar D/k_B T}$ is the thermal diffusion length. When $L_T \gg \xi$, $f(\xi/L_T) = A + B(\xi/L_T)$, which is equivalent to Eq. (3). In the critical region, where $L_T \ll \xi \to \infty$, Eq. (6) should be reduced to

$$\sigma = C \frac{e^2}{\hbar L_T} . \qquad (7)$$

Combining this equation and Eq. (5), they obtained $\sigma \propto T^{1/3}$. More recently, the $T^{1/3}$ dependence has been predicted based solely on the effect of disorder[41] and on the quantum interference.[42]

Although the origin for the $T^{1/3}$ dependence in the present system near $N_c$ is unknown at this point, it is important that we find a method that allows the determination of $\sigma(0)$ even when the temperature dependence of $\sigma(T)$ changes from $\sqrt{T}$ to $T^{1/3}$ as $N$ approaches $N_c$. For this purpose we follow Al'tshuler and Aronov's manipulation[38] of eliminating $m$ and $D$ in Eqs. (3)–(5) and obtain

$$\sigma(T) = \sigma(0) + m'\sqrt{T/\sigma(T)} , \qquad (8)$$

where $m' = Ae\sqrt{(\partial n/\partial \mu)}$, which is temperature independent. In the limit of $\Delta\sigma(T) \ll \sigma(0) \approx \sigma(T)$, this equation gives the same value of $\sigma(0)$ as Eq. (3) does. When $\sigma(0) \ll \sigma(T)$, it yields a $T^{1/3}$ dependence for $\sigma(T)$. Thus it is applicable to both $\sqrt{T}$ and $T^{1/3}$ dependent conductivity. From today's theoretical understanding of the problem, Eqs. (3) and (8) are valid only for $L_T \gg \xi$, and their applicability to the critical region is not clear, because the higher-order terms of the $\beta$ function[4] which were once erroneously believed to be zero do not vanish.[39,40] Nevertheless, we expect Eq. (8) to be a good expression for describing the temperature dependence of all metallic samples because it expresses both the $\sqrt{T}$ and the $T^{1/3}$ dependences as limiting forms. Then, based on Eq. (8) we plot $\sigma(T)$ vs $\sqrt{T/\sigma(T)}$ for the four close to $N_c$ samples in Fig. 3. As we see, the data points align on straight lines very well, which supports the adequacy of Eq. (8). The zero-temperature conductivity $\sigma(0)$ is obtained by extrapolating to $T = 0$. The curve on the top of Fig. 3 is for the sample with the lowest $N$ among the ones showing $\sqrt{T}$ dependence at low temperatures, i.e., this sample has the largest value of $\Delta\sigma(T)/\sigma(0)$ among $\sqrt{T}$ samples. The value of $\sigma(0)$ obtained for this particular sample using Eq. (8) differs only by 0.6% from the value determined by the conventional extrapolation assuming Eq. (3). This small difference is comparable to the error arising from the choice of the temperature range in which the fitting is performed. Therefore the new extrapolation method proposed here is compatible with the conventional method based on the $\propto \sqrt{T}$ extrapolation. A different method for the determination of $\sigma(0)$ was proposed recently by Shlimak *et al.*[31] but it requires $\sigma(T) = a + bT^{1/3}$ for all samples with exactly the same $b$. Such a strict condition is not met in Ge:Ga and in many other systems as we will show later in Fig. 5.

Based on our new analysis, the MI transition was found to occur between the first and second samples from the bottom in Fig. 3. Here $N_c$ is fixed already within an accuracy of 0.16% corresponding to the fractional difference in $N$ between the first and second samples from the bottom, i.e., unlike the case for Si:P,[30,43,44] the determination of the critical conductivity exponent will *not* be affected by the ambiguity in the value of $N_c$. Figure 4 shows the $\sigma(0)$ as a function of $N/N_c - 1$ with an excellent fit by Eq. (1) (dotted line) with $\nu = 0.50 \pm 0.04$ and $N_c = 1.860 \times 10^{17}$ cm$^{-3}$ all the way down to $(N/N_c - 1) = 4 \times 10^{-4}$. The clear demonstration of the same $\nu \approx 0.50$ in our previous work[19] was criticized by Shlimak *et al.*[45] for our doping level's not being close enough to $N_c$. The present work shows that the critical exponent is indeed $\approx 0.5$ for nominally uncompensated Ge:Ga. We note that $\nu = 0.46 \pm 0.18 \approx 0.5$ is obtained even when fitting only the results obtained with the four samples closest to the transition. As was mentioned in the Introduction, Eq. (1) with $\nu \approx 0.5$ holds for many nominally uncompensated crystalline semiconductors for a relatively wide range of $N$ above $N_c$. The $\sigma(0)$ of Ge:Ga shown in Fig. 4 can be fitted very well with a single exponent $\nu \approx 0.50$ over three orders of magnitude in $N/N_c - 1$. In fact it was shown in Ref. 19 that $\nu \approx 0.5$ holds up to $N = 1.4N_c$.

In order to compare the low-temperature transport properties of Ge:Ga with other systems, we evaluate the concentration $N^*$ where the sign of $d\sigma/dT$ changes. $N^*$ of our $p$-type Ge:Ga lies between $1.04N_c$ and $1.08N_c$, while larger values of $1.2N_c < N^* < 1.3N_c$ have been reported for $n$-type germanium; Ge:Sb (Ref. 46) and Ge:As.[31] The magnitude of $\Delta\sigma(T)$ in Ge:Ga is considerably smaller than that of Ge:Sb (Ref. 46) for samples with approximately the same $N/N_c - 1$. A number of properties related to the band structure, e.g., the valley degeneracy, strength of the spin-orbit interaction, the degree of the intervalley scattering, etc., can change the low-temperature transport properties of doped semiconductors. The difference in the behavior of $\sigma(T)$ at finite temperature between $p$- and $n$-type Ge may be understood in such contexts. Concerning the critical behavior of $\sigma(0)$ at the MI transition, it is usually thought to depend on the universality class to which the system belongs, and can vary depending on the strength of the spin-orbit scattering or of the spin scattering. From the experiments so far done, including the present one on doped semiconductors except $n$-type Ge, we conclude, however, that the critical exponent $\nu \approx 0.5$ applies, irrespective of the systems as long as the compensation is not important. Regarding $n$-type Ge, $\nu \approx 1$ was reported in Ge:As (Ref. 31) and Ge:Sb (Ref. 31) and $\nu \approx 0.9$ in Ge:Sb.[46] In order to verify whether this is truly the case, an investigation of $n$-type NTD $^{74}$Ge:As is important. It is also interesting to point out that $N^*$ of Ge:Ga is very similar to $N^*$ found in both $p$- and $n$-type Si. The $N^*$ for Si:B (Ref. 47) is about $1.08N_c$ and for Si:P (Refs. 30 and 48) lies between $1.03N_c$ and $1.2N_c$. The coefficient $m$ in Eq. (3) is compared in Fig. 5 for Ge:Ga, Si:P, and Si:B systems.

### B. Variable range hopping conduction in insulating samples

The temperature dependence of the resistivity of insulating samples is shown in Fig. 6. Shklovskii and Efros have shown for insulating samples that a parabolic shaped energy gap (known as the Coulomb gap) exists in the sigle-particle density of states in the immediate vicinity of the Fermi level.[49] The variable range hopping resistivity $\rho$ for the excitation within the Coulomb gap is given by[49]

$$\rho = \rho_0 \exp[(T_0/T)^p], \tag{9}$$

where $\rho_0$ is a prefactor, $p = 1/2$, and

$$k_B T_0 \approx \frac{1}{4\pi\epsilon_0} \frac{2.8 e^2}{\kappa(N)\,\xi(N)}. \tag{10}$$

$\kappa(N)$ and $\xi(N)$ are the dielectric constant and localization length, respectively. Moreover, $\kappa(N) = \kappa_0(1 - N/N_c)^{-s}$ and $\xi(N) = \xi_0(1 - N/N_c)^{-\zeta}$ as $N$ approaches $N_c$ from the insulating side so that $T_0$ becomes

$$k_B T_0 \approx \frac{2.8 e^2}{4\pi\epsilon_0 \kappa_0 \xi_0} (1 - N/N_c)^\alpha, \tag{11}$$

where $\alpha = s + \zeta$ is to be determined experimentally in this study. Because the width of the Coulomb gap $\Delta_{\rm CG}$ depends also on $N$ via $\kappa(N)$ as $\Delta_{\rm CG} \propto [\kappa(N)]^{-3/2}$, it collapses rapidly as $N$ approaches to $N_c$ from the insulating side. When $\Delta_{\rm CG}$ becomes sufficiently small near $N_c$, the excitation energy for hopping given by the thermal energy can become larger than $\Delta_{\rm CG}$.[50] In this case the density of states may be considered to be constant around the Fermi level and the Mott variable range hopping with $p = 1/4$ in Eq. (9) is expected to be observed. Such a crossover from $p = 1/2$ to $p = 1/4$ as $N$ approaches $N_c$ was observed in Si:P.[51] It is of great interest to see if such a crossover exists in our homogeneously doped Ge:Ga system. Figure 7 shows the values of $p$ found from the calculation of $d\ln\varepsilon/d\ln T$ where $\varepsilon \equiv -d\ln\rho/d\ln T$ and/or from the direct fitting of curves shown in Fig. 6 by Eq. (9). $p \approx 1/2$ is obtained for the samples having $N < 0.991N_c$, i.e., they will be analyzed in the framework of Shklovskii and Efros's theory for the hopping within the Coulomb gap.[49] In Figs. 8(a) and 8(b), $T_0$ and $\rho_0$, respectively, are plotted as a function of $(1 - N/N_c)$ for the samples with $N < 0.991N_c$. As was already shown in our previous work,[19] the best fit of $T_0$ to Eq. (11) is obtained here with $\alpha \approx 1$. Based on this finding, we calculate $T_0$ using Eq. (11) with $\alpha = 1$, $\kappa_0 = \epsilon = 16$, and $\xi_0 = 4\pi\epsilon_0\epsilon\hbar^2/m^*e^2 = 8$ nm, where $\epsilon$ is the dielectric constant of Ge and $m^*$ is the effective mass of the electron in Ge.[52] The calculated (not fitted) $T_0$, which is shown by the dotted line in Fig. 8(a), agrees very well with the experimentally determined $T_0$, supporting the quantitative validity of the theoretical expression for $T_0$.

In some of the earlier studies, the constant 2.8 in Eq. (10) had to be adjusted to much smaller values in order to obtain an agreement with experimentally found $T_0$.[53,54] In Fig. 8(b), $\rho_0$ is shown as a function of $1 - N/N_c$.

The prefactor $\rho_0$ shows no critical behavior and it approaches near $N_c$ a value very close to the inverse of Mott's minimum metallic conductivity denoted by the dotted line. Finally we turn our attention to $p$ of the samples having $0.991N_c < N < N_c$ in Fig. 7. In this regime lying very close to $N_c$, $1/p$ increases rapidly as $N$ approaches $N_c$ due to the collapsing of the Coulomb gap. However, $1/p$ does not approach a constant value of 4 expected for the Mott variable range hopping conduction. In our analysis the temperature dependence of the prefactor $\rho_0$, which can be significant near $N_c$, is neglected. Therefore further analysis taking into account the appropriate dependencies of $\rho_0$ on $T$ is important. Unfortunately we cannot perform such an analysis with the accuracy needed at this point since the theoretical models proposed so far on $\rho_0$ do not agree with one another.[55]

## IV. CONCLUSION

We have measured the electrical conductivity of nominally uncompensated neutron-transmutation-doped isotopically enriched $^{70}$Ge:Ga samples. Approaching the transition from the metallic side, we find that the temperature dependence of the form $\sigma(T) = a + bT^q$ with $q = 1/2$ is replaced by $q = 1/3$. We introduce a method for finding $\sigma(0)$ which is consistent with the conventional $\sqrt{T}$ extrapolation. The critical conductivity exponent $\nu \approx 0.5$ for $p$-type germanium has been fully confirmed. On the insulating side of the MI transition, the standard relation for the variable range hopping resistivity $\rho(T) \propto \exp(T_0/T)^p$ with $p = 1/2$ is observed for $N < 0.991N_c$. Shklovskii and Efros's expression for $T_0$ agrees quantitatively with our experimentally found $T_0$.

## ACKNOWLEDGMENTS

We are thankful to T. Ohtsuki for valuable discussions, J. W. Farmer for the neutron irradiation, and V. I. Ozhogin for the supply of the Ge isotope. The valuable comments by J. C. Phillips were greatly appreciated. The work at Keio was supported by the Kurata Foundation, the Iketani Science and Technology Foundation, and a Grant-in-Aid for Scientific Research from the Ministry of Education, Science, Sports, and Culture, Japan. The work at Berkeley was supported in part by the Director, Office of Energy Research, Office of Basic Energy Science, Materials Sciences Division of the U. S. Department of Energy under Contract No. DE-AC03-76SF00098 and in part by U. S. NSF Grant No. DMR-94 17763.

FIG. 1. Electrical conductivity as a function of $T^{1/2}$ for NTD $^{70}$Ge:Ga. From bottom to top in units of $10^{17}$ cm$^{-3}$, the concentrations for the samples denoted by solid symbols are 1.853, 1.856, 1.858, 1.861, 1.863, 1.912, 2.210, and 2.232, respectively. Open symbols are the data taken on the samples used in our previous work (Ref. 19).

FIG. 2. Conductivity as a function of (a) $T^{1/2}$ and (b) $T^{1/3}$, respectively, near the MI transition. From bottom to top in units of $10^{17}$ cm$^{-3}$, the concentrations are 1.853, 1.856, 1.858, 1.861, 1.863, and 1.912, respectively. The upper and lower dotted lines in each figure represent the best fit using the data between 0.05 K and 0.5 K for the first and the third curves from the top, respectively. Each fit is shifted downward slightly for easier comparison.

FIG. 3. Conductivity $\sigma$ as a function of $(T/\sigma)^{1/2}$. From bottom to top in units of $10^{17}$ cm$^{-3}$, the concentrations are 1.858, 1.861, 1.863, and 1.912, respectively. The solid lines denote the extrapolation for finding $\sigma(0)$.

FIG. 4. Zero-temperature conductivity $\sigma(0)$ vs the dimensionless distance $N/N_c - 1$ from the critical point on a double logarithmic scale. The dotted line represents the best power-law fit by $\sigma(0) \propto (N/N_c - 1)^\nu$ where $\nu = 0.50 \pm 0.04$. The open symbols are from our previous work (Ref. 19).

FIG. 5. Coefficient $m$ defined in Eq. (3) as a function of $N/N_c - 1$; Ge:Ga of this work ($\bullet$), Ge:Ga of the previous work (Ref. 19) ($\circ$), Si:B (Ref. 47) ($\triangle$), and Si:P (Ref. 48) ($\square$).

FIG. 6. The logarithm of the resistivity as a function of $T^{-1/2}$ for insulating samples. The triangles denotes the data from Ref. 19 with the fit by Eq. (9) (solid line). The samples of the present study are represented by circles and the concentrations from top to bottom in units of $10^{17}$ cm$^{-3}$ are 1.840, 1.842, 1.843, 1.848, 1.850, 1.853, 1.856, and 1.858, respectively.

FIG. 7. The inverse of the exponent $p$ defined by Eq. (9) vs concentration. The open circles are from Ref. 19.

FIG. 8. On a double logarithmic scale, $T_0$ and $\rho_0$ are plotted as functions of $1 - N/N_c$ in (a) and (b), respectively. The open symbols are after Ref. 19. The dotted line in (a) is a calculated $T_0$ using Eq. (11) with $\alpha = 1$. The dotted line in (b) represents the inverse of the Mott minimum metallic conductivity.

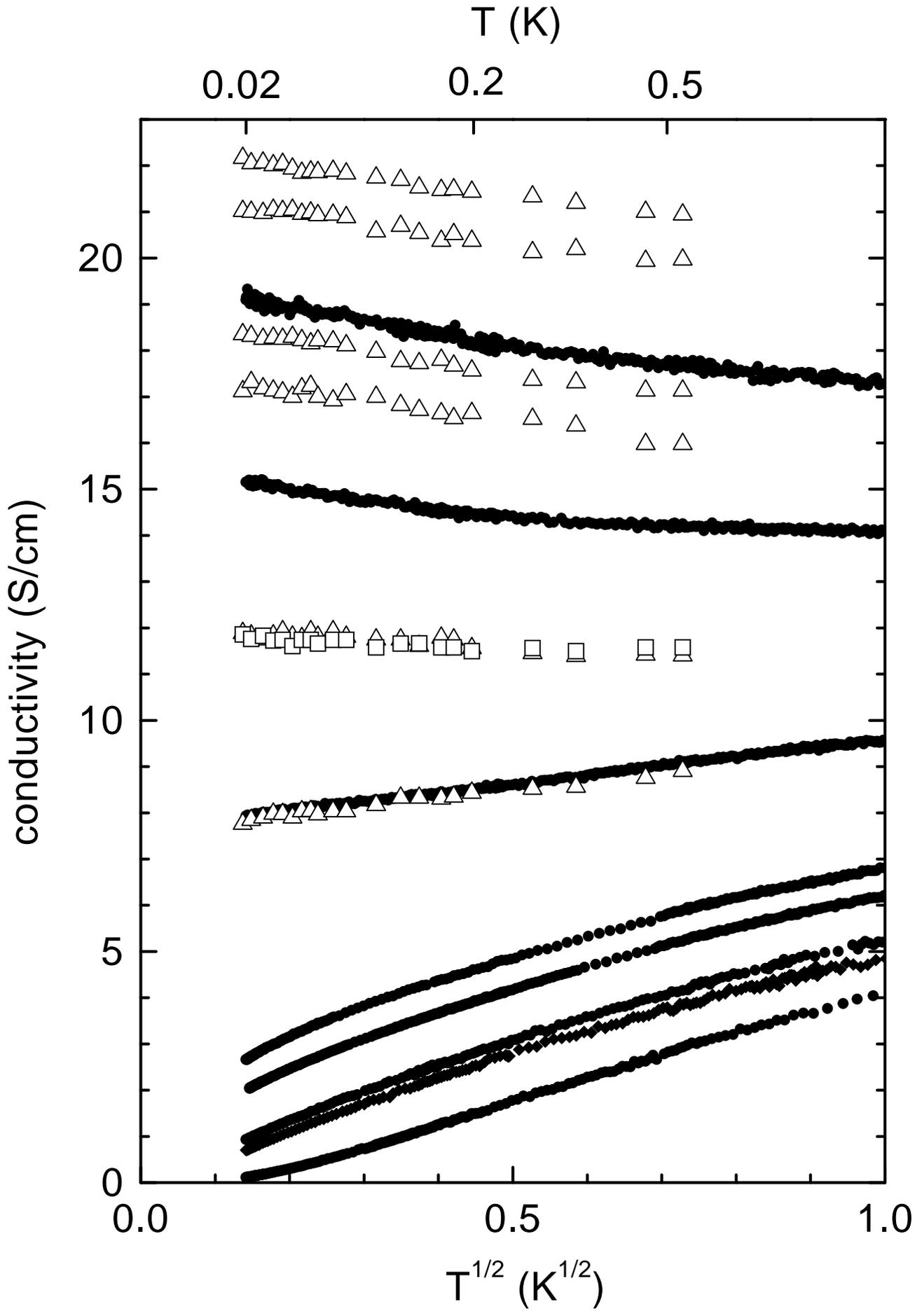

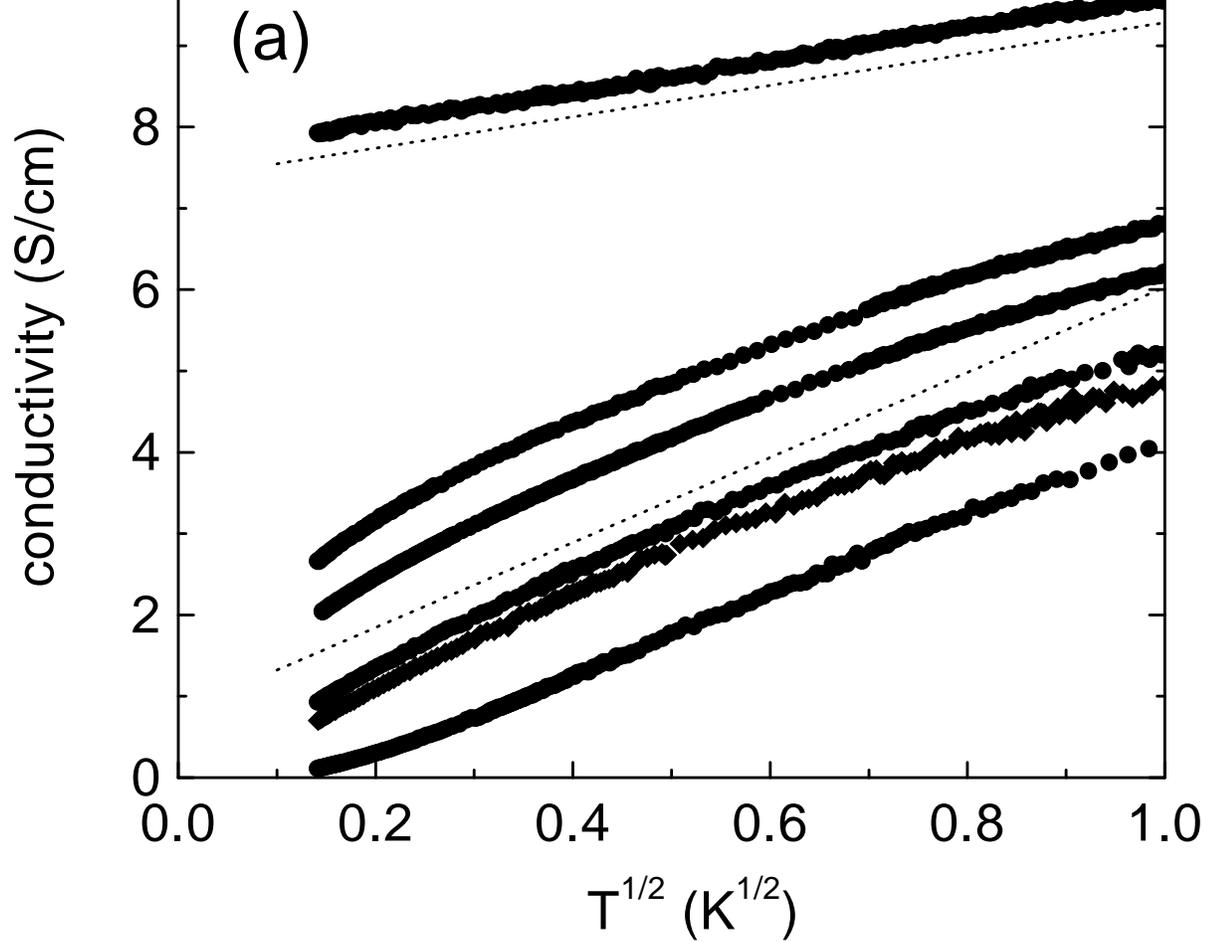

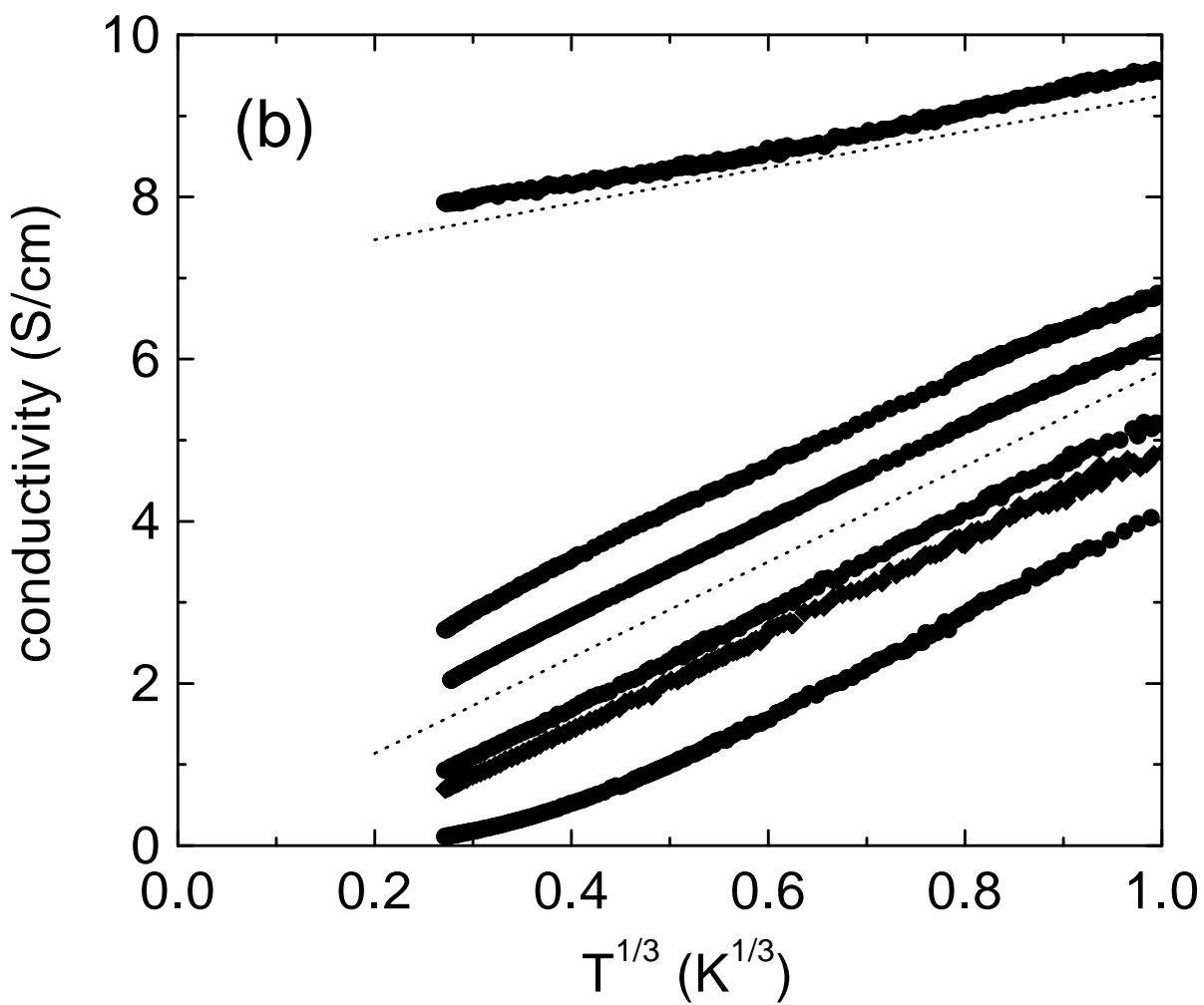

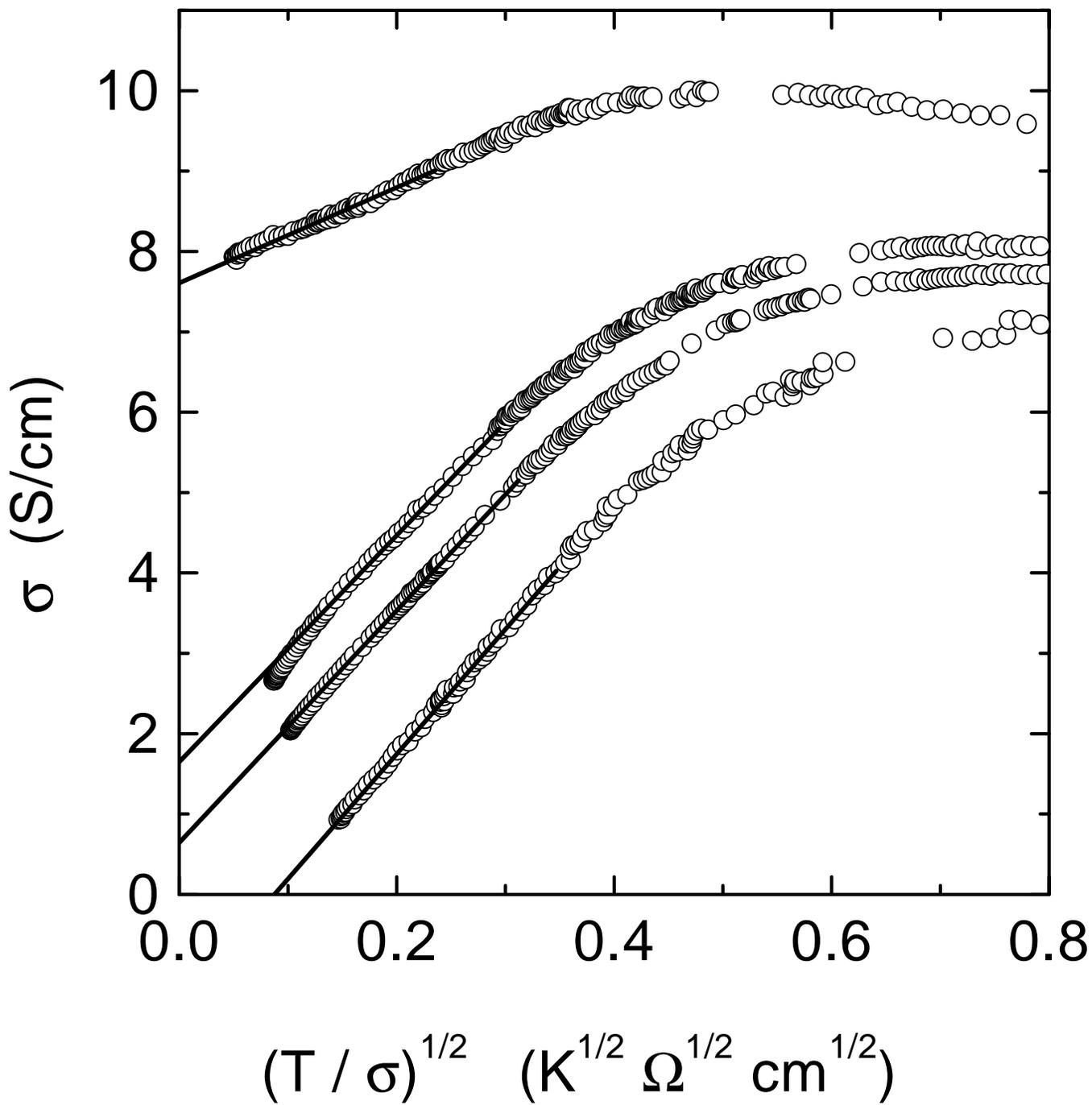

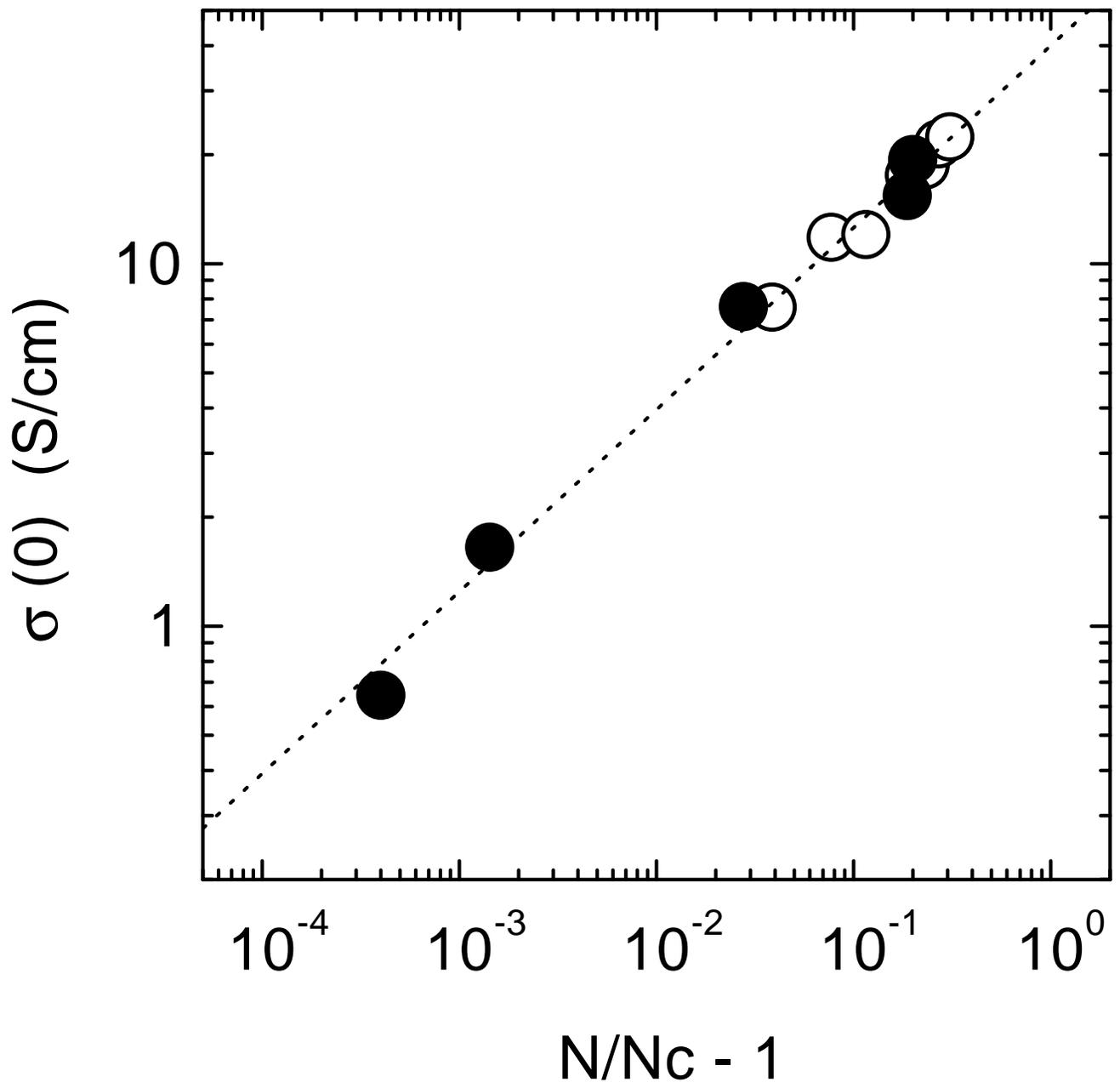

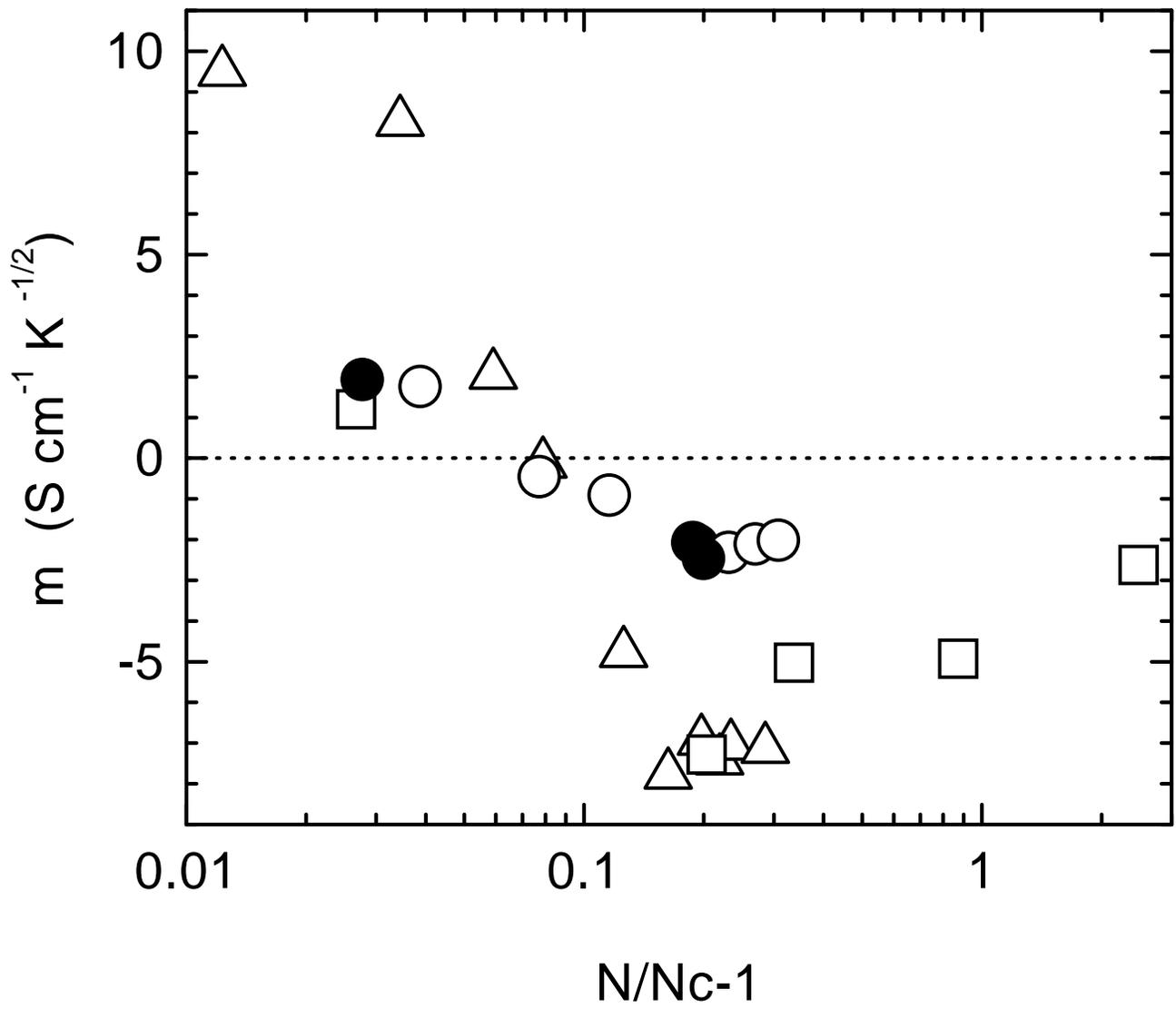

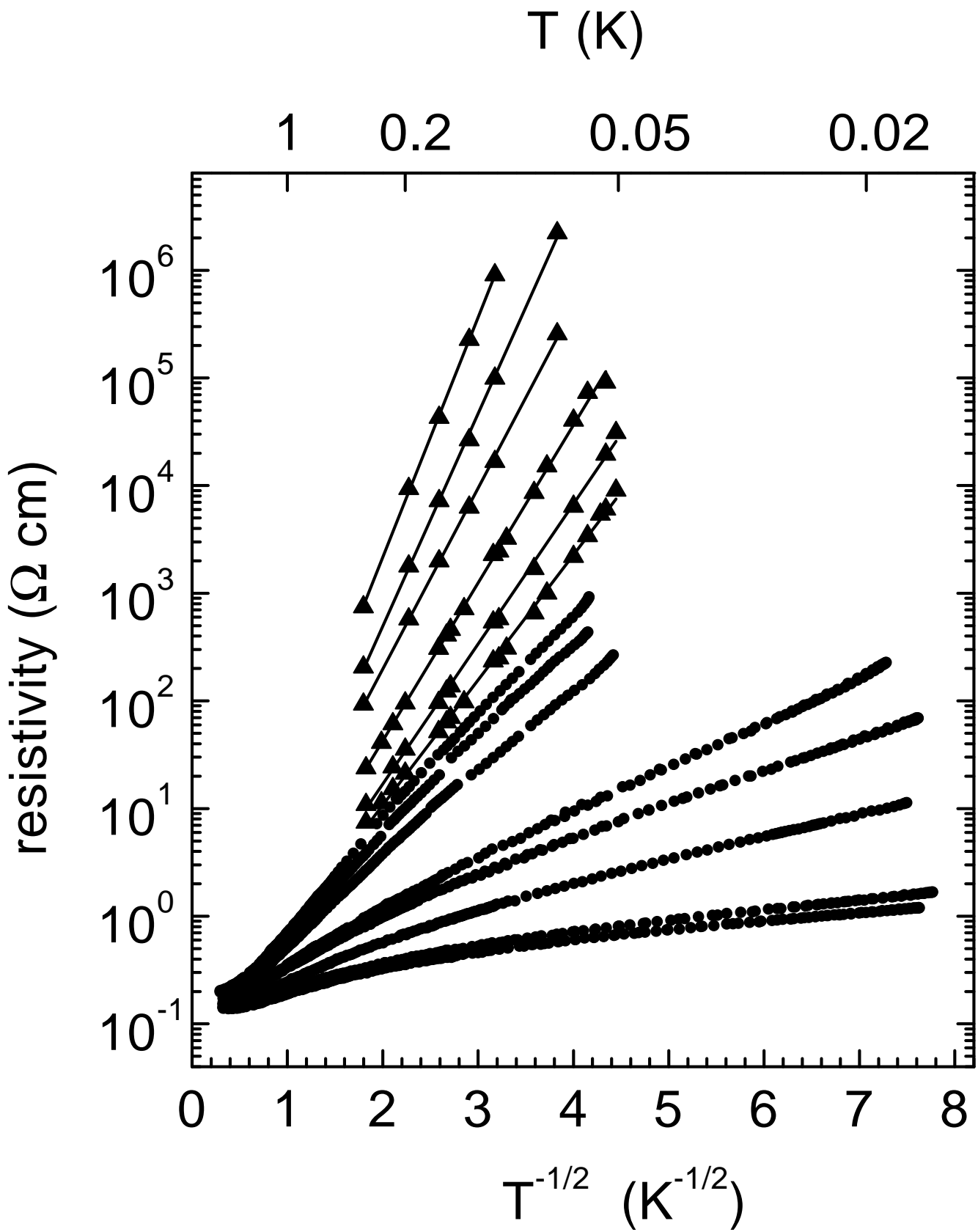

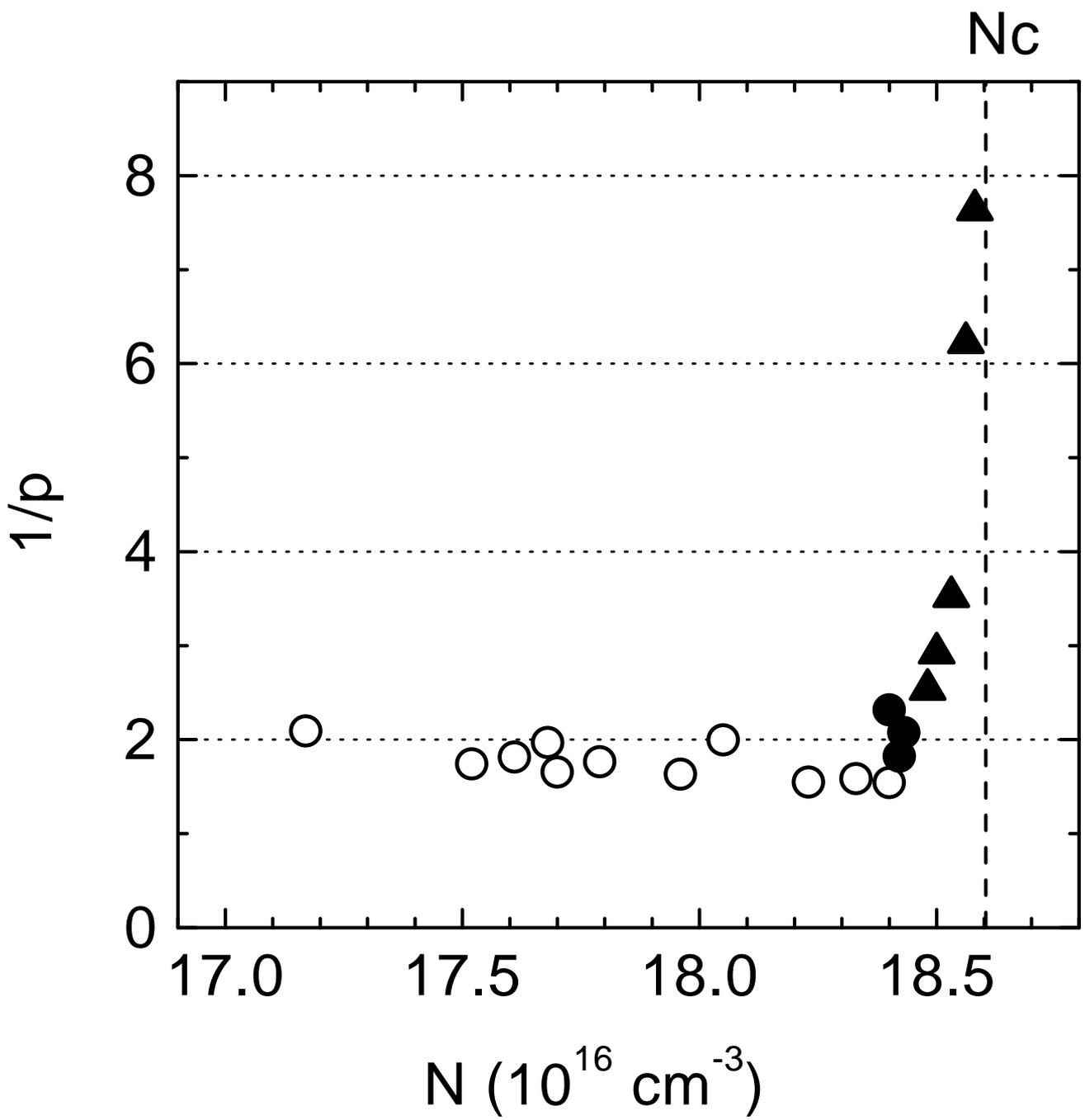

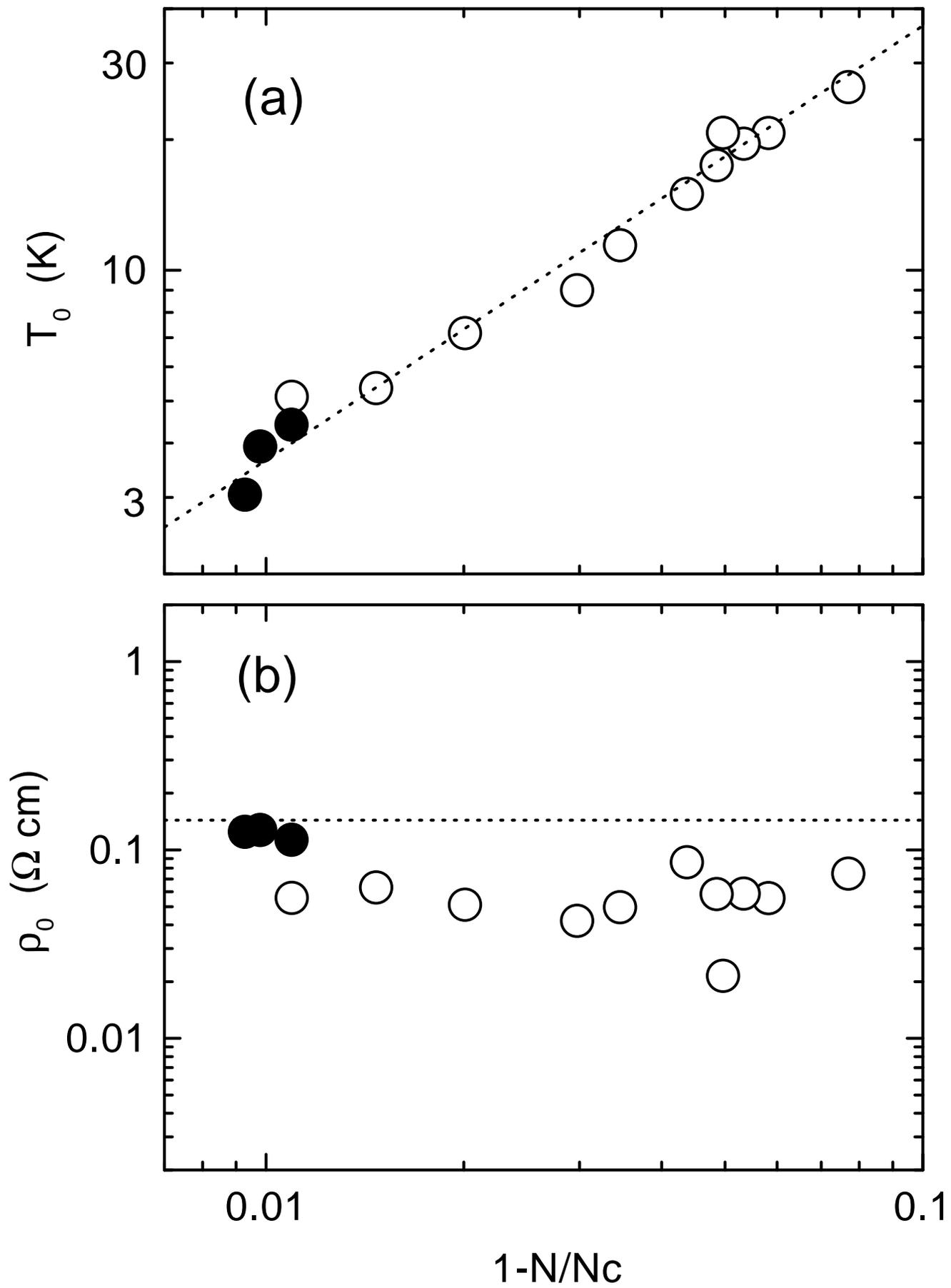